\documentclass[conference]{IEEEtran}
\usepackage{setspace}
\usepackage{clrscode}
\usepackage{amsmath}
\usepackage{amssymb}
\usepackage[dvips]{graphicx}
\usepackage{epsfig}
\usepackage{hyperref}
\usepackage{bm}
\usepackage{subfigure}
\usepackage{verbatim}
\usepackage{titletoc}
\usepackage{extarrows}
\usepackage{fancyhdr}

\usepackage{balance}

\newcommand{\C}{{\sf{C}}}

\newcommand{\figsize}{0.45}

\newtheorem{Lem1}{Proposition}

\newtheorem{Rem}{Remark}

\newtheorem{Def}{Definition}
\begin{document}

\title{Alternate Distributed Beamforming for Decode-and-Forward Multi-Relay Systems Using Buffers}
    \author{
\IEEEauthorblockN{Jiayu Zhou and Deli Qiao}
\IEEEauthorblockA{\small{School of Information Science and Technology,}}
\small{East China Normal University, Shanghai, China}\\
\small{Email: 51171214021@stu.ecnu.edu.cn, dlqiao@ce.ecnu.edu.cn}}

\maketitle

\begin{abstract}\footnote{This work has been supported in part by the National Natural Science Foundation of China (61671205).}
In this paper, link selection for half-duplex buffer-aided relay systems is revisited. A new fixed scheduling policy referred as alternate distributed beamforming (ADB) is proposed, in which the relays are divided into two groups, with one group receiving the same information broadcast from the source and the other group transmitting the common messages to the destination via distributed beamforming in each time slot. It is worth noting that the relays used for reception and transmission are determined without the need of instantaneous channel state information (CSI). Theoretical analysis of the achievable throughput of the proposed scheme in Rayleigh fading is provided and the approximate closed-form expressions are derived. Through numerical results, it is shown that compared with existing link selection policies, the proposed fixed scheduling ADB achieves a significant improvement in achievable throughput.
\end{abstract}

\section{Introduction}
Cooperative communications have been an important building block for communication systems, in which the communication between a source node and a destination node is accomplished via the help of a number of relay nodes
\cite{cooperative communication}. With the help of relays, alternative and independent transmission paths are offered, and the diversity gain of the network can be obtained. Also, distributed beamforming gain can be expected \cite{DF}.

To better utilize the benefits provided by multiple relays, various relay selection schemes have been proposed. The conventional relay selection (CRS) scheme selects single relay that provides the strongest end-to-end path between the source and destination \cite{CRS}. The selected relay forwards the received data immediately in the next time slot to the destination and therefore the ability of the relays to store at least a limited number of data packets is not elaborated. This relay selection policy is considered as the optimal selection scheme for conventional relaying system without buffers. The adoption of buffer-aided relays can provide both throughput and diversity gain by adaptive link selection \cite{buffer1}, \cite{buffer2}. Storing packets and transmitting them in favorable wireless conditions increases the network's resiliency, throughput and diversity. A max-max relay selection (MMRS) scheme was proposed in \cite{max-max}, which selects the relay with the best source-relay link and the best relay-destination link for data reception and transmission, respectively. And space full-duplex max-max relay selection (SFD-MMRS) scheme was introduced in \cite{mimick}, which mimics full-duplex (FD) relaying with half-duplex (HD) relays via link selection. A new relay selection scheme called max-link relay selection scheme was suggested in \cite{max-link}, which selects the strongest link for transmission among all the available links at each time slot. In addition, a modified max-link relay selection scheme has also been proposed, in which the direct link between the source and destination was exploited to achieve a significant performance gain in terms of diversity and delay \cite{modified max-link}. And a general relay selection factor including the weight of the link and the link quality was defined in \cite{weight}. It is worth noting that the relay selected with the adaptive link selection policies varies with the instantaneous channel state information (CSI), which may introduce more complexities in practical implement.

Note that a relay usually operates in either FD or HD mode. In FD relaying, the relays transmit and receive at the same time, and this operation will increase the hardware complexity \cite{2}, \cite{3}. We consider HD relays in this paper. In HD relaying, relays are incapable of transmitting and receiving simultaneously, thus leading to reduced capacity of the whole network. In order to recover the HD loss, several successive relaying schemes have been proposed \cite{spectral efficient}-\cite{bypassing}, the main idea of which is to adopt two relays acting as the receiving and transmitting relay simultaneously. In \cite{spectral efficient}, a two-way relaying protocol and a two-path relaying protocol were proposed. It was shown that both protocols recover a significant portion of the HD loss for different relaying strategies. Both inter-relay interference (IRI) and a direct link between source and destination were considered in \cite{recover loss}. In \cite{bypassing}, the authors considered a two-relay network with multiple antennas at the destination. The infrastructure-based relays with highly directional antennas were used to avoid IRI. Note however that a two-relay network is assumed in most of the aforementioned works with successive relaying.

Inspired by the decode-and-forward \cite{DF} and fixed scheduling \cite{1}, in this paper, we propose a novel relaying protocol named alternate distributed beamforming (ADB), where the receiving and transmitting relays are predetermined in each time slot by grouping the relays. We consider the half-duplex buffer-aided multiple-relay systems. We assume that there is no direct link between the source and the destination. We analyze the achievable throughput of the proposed scheme in Rayleigh fading channels and derive the closed-form expressions. Numerical results in accordance with theoretical analysis show the superiority of ADB.

The reminder of this paper is organized as follows. The system model and several existing protocols are briefly introduced in Section II. In Section III, the operation of ADB is described in detail, and comprehensive analysis of the achievable throughput is presented and the approximate closed-form expressions are derived. Numerical results are provided in Section IV. Finally, conclusions are drawn in Section V with some lengthy proofs in Appendix.

\section{Preliminaries}
\begin{figure}
    \centering
    \includegraphics[width=0.3\textwidth]{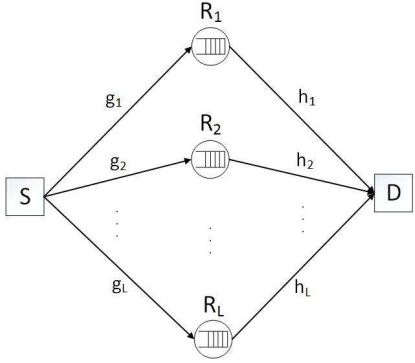}
    \caption{System model.}
    \label{fig:figure1}
\end{figure}
\subsection{System Model}
We consider a relay network consisting of one source node $S$, a set of $L$ decode-and-forward (DF) relays $R_{1},...,R_{L}$, and one destination node $D$, as shown in Fig. \ref{fig:figure1}. We assume that all nodes are equipped with a single antenna and operate in the HD mode, i.e., they cannot transmit and receive data simultaneously. We assume that there is a buffer of infinite length at each relay such that each relay can store the information received from the source and transmit it in later time.

We assume that there is no direct link between the source and destination, and the communications can be established only via relays. We use $g_{i}$ and $h_{i}$ for $i=1,...,L$ to denote the channel coefficients of $S-R_i$ and $R_{i}-D$ links, respectively. The channel is assumed to be stationary and ergodic. We consider the block fading, in which the channel coefficients remain constant during one time slot and vary independently from one to the other. In addition to fading, all wireless links are impaired by additive white Gaussian noise (AWGN). The instantaneous received signal-to-noise ratio (SNR) at relay $R_i$ is given by $\gamma_{g_{i}}=\frac{P_{S}|g_{i}|^2}{N_{0}}$, and the instantaneous received SNR at the destination from relay $R_i$ is given by $\gamma_{h_{i}}=\frac{P_{R}|h_{i}|^2}{N_{0}}$, where $P_S$ is the transmit power of the source, $P_R$ is the transmit power of each relay, and $N_0$ is the noise power at the relays and destination.

We assume Rayleigh fading for the channel coefficients and the variances of $g_{i}$ and $h_{i}$ are assumed to be $\sigma_{g_{i}}^2$ and $\sigma_{h_{i}}^2$, respectively. Throughout this paper, we consider the case of independent and identically distributed (i.i.d.) fading for both $S-R_i$ and the $R_{i}-D$ links, i.e. $\sigma_{g_{i}}^2=\sigma_{g}^{2}$ and $\sigma_{h_{i}}^2=\sigma_{h}^{2}, i=1,...,L$, which is a typical assumption to facilitate analysis \cite{mimick}.
\subsection{Existing Relaying Protocols}
In this part, we review several existing relaying protocols. It is assumed that CSI is known at the transmitter of each link.

\subsubsection{Conventional Relay Selection (CRS)}
The conventional relay selection protocol selects the relay which provides the strongest end-to-end path between the source and destination \cite{CRS}. The source transmits in the first time slot and the selected relay forwards the data received from the source towards the destination in the second time slot. The best relay $R_j$ is selected based on
\begin{align}
j= {\rm arg} \max\limits_{i\in\{1,...,L\}}\{{\rm min}\{\gamma_{g_{i}},\gamma_{h_{i}}\}\}.
\end{align}
The instantaneous end-to-end capacity for the overall system is given by
\begin{align}
C_{k}=\frac{1}{2}{\rm log_{2}}\left(1+{\rm \max\limits_{1\leq{k}\leq{L}}}{\rm min}\left(P_{S}|g_{k}|^{2}, P_{R}|h_{k}|^{2}\right)\right).
\end{align}
Then, the achievable throughput is given by ${\mathbb E}[C_{k}]$, where ${\mathbb E}[\cdot]$ denotes the expectation. Throughout this text, the unit for the throughput is bps/Hz.

\subsubsection{Space Full-Duplex Max-Max Relay Selection (SFD-MMRS)}
This protocol chooses different relays for reception and transmission, according to the quality of the channels, so that the relay selected for reception and the relay selected for transmission can receive and transmit at the same time \cite{mimick}. The best relay for reception $R_{r_{1}}$ and the best relay for transmission $R_{t_{1}}$ are selected respectively based on
\begin{align}
r_1={\rm arg} \max\limits_{i\in\{1,...,L\}}\{\gamma_{g_{i}}\},
\end{align}
\begin{align}
t_1={\rm arg} \max\limits_{i\in\{1,...,L\}}\{\gamma_{h_{i}}\}.
\end{align}
The second best relay for reception $R_{r_{2}}$ and the second best relay for transmission $R_{t_{2}}$ are selected respectively according to
\begin{align}
r_2=\arg \max_{\underset{i\neq r_1}{i\in \{1, \ldots, L\}}}  \{\gamma_{g_i}\},
\end{align}
\begin{align}
t_2=\arg \max_{\underset{i\neq t_1}{i\in \{1, \ldots, L\}}}  \{\gamma_{h_i}\}.
\end{align}
Then, in SFD-MMRS, the relays selected for reception $R_{\bar{r}_1}$ and transmission $R_{\bar{t}_1}$ are chosen as
\begin{align}
(R_{\bar{r}_1}, R_{\bar{t}_1})=
\begin{cases}
(R_{r_1}, R_{t_1}), \text{if} \quad r_1\neq t_1\\
(R_{r_2}, R_{t_1}), \text{if} \quad r_1= t_1 \text{and} \min(\gamma_{g_{r_2}}, \gamma_{h_{t_1}})\\
                         \hspace {1.8cm}   > \min(\gamma_{g_{r_1}}, \gamma_{h_{t_2}})\\
(R_{r_1}, R_{t_2}),  \text{otherwise}.
\end{cases}
\end{align}
Let $C_{SR}$ and $C_{RD}$ denote the instantaneous capacities of the $S-R$ and $R-D$ links, respectively, i.e.,
\begin{align}
C_{SR}&={\rm log_{2}}\left(1+P_{S}|g_{\bar{r}_1}|^{2}\right),\ \nonumber\\
C_{RD}&={\rm log_{2}}\left(1+P_{R}|h_{\bar{t}_1}|^{2}\right).
\end{align}
The achievable throughput is given by ${\rm min}\{{\mathbb E}[C_{SR}], {\mathbb E}[C_{RD}]\}$.

\subsubsection{Decode and Forward (DF)}
In DF \cite{DF}, each relay must decode the common message transmitted by the source node and beamform their transmissions to the destination, which is also performed in two time slots. Then, the instantaneous rate for the overall system is given by
\begin{small}
\begin{align}
C_{k}=\frac{1}{2}{\rm log_{2}}\left(1+{\rm min}\left(\min\limits_{1\leq{k}\leq{L}}|g_{k}|^{2},\left(\sum\limits_{k=1}^{N}|h_{k}|\right)^{2}\right)\right).
\end{align}
\end{small}
The achievable throughput is given by ${\mathbb E}[C_{k}]$.

\section{Alternating Decode-and-Forward Protocol}

\subsection{The transmission policy}
\begin{figure}
    \centering
    \includegraphics[width=\figsize\textwidth]{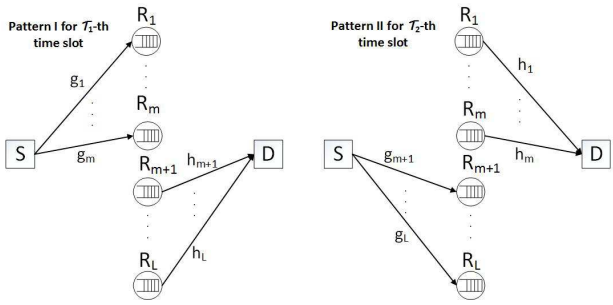}
    \caption{Transmission modes of the proposed scheme.}
    \label{fig:figure2}
\end{figure}
The operation of the ADB can be seen in Fig. \ref{fig:figure2}, which has two patterns. Time is slotted into discrete equal-size time slots. We divide $L$ relays into two groups, i.e., group 1 with $m$ relays, ${\mathcal R_{1}}=\{R_{1},...,R_{m}\}$ and group 2 with $L-m$ relays, ${\mathcal R_{2}}=\{R_{m+1},...,R_{L}\}$ \footnote{Due to the i.i.d. assumption, the relays can be divided arbitrarily. In case of different fading statistics, the relay grouping will be another interesting problem.}. The source broadcasts messages to the relays in group ${\mathcal R_{1}}$ for each ${t_{1}}$-th time slot while at the same time, the relays in group ${\mathcal R_{2}}$ beamform the data available in their buffers to the destination. It is assumed that the relays are synchronized through signaling. Similarly, during the ${t_{2}}$-th time slot, the relays in group ${\mathcal R_{2}}$ must decode the message transmitted by the source node and stores the packet in their buffers while the relays in group ${\mathcal R_{1}}$ beamform the previously received packets to the destination. Denote ${\mathcal T_{1}}=\{t_1\}$, ${\mathcal T_{2}}=\{t_2\}$ as the set of time indices for the relays in group 1 and 2 receiving data from the source, respectively. Note that ${\mathcal T_{1}} \cup {\mathcal T_{2}}=\{1,2,...,N\}$, where $N$ is the total number of time slots. We assume that the cardinality of ${\mathcal T_{1}}$ and ${\mathcal T_{2}}$ is $|{\mathcal T_{1}}|=|{\mathcal T_{2}}|=\frac{N}{2}$.

In this strategy, the benefits of both DF and fixed scheduling are enjoyed. It is obvious that with this protocol, the HD loss of conventional relays can be recovered \cite{mimick} and distributed beamforming gain can be expected. And it is worth noting that compared with the selective protocols CRS and SFD-MMRS, the receiving and transmitting relays in the proposed policy do not vary with the instantaneous CSI and are predetermined at the beginning of transmissions, which makes it easier to implement in practice.

\subsection{Achievable Throughput Analysis}
In this section, we analyze the achievable throughput performance of the proposed ADB scheme and derive the approximate closed-form expressions. Due to the assumption of no inter-relay links \cite{mimick}, \cite{bypassing}, we assume that there is no inter-relay interference when the receiving relays and transmitting relays are active in the same time-slot. In practice, this assumption is valid if the relays are located far away from each other or if fixed infrastructure-based relays with directional antennas are used. Note that fixed relays are of practical interest since they are low-cost and low-transmit power devices (see, e.g., \cite{bypassing}, \cite{frelay2}, and \cite{frelay3}). Without loss of generality, we assume that the noise power at the receiving nodes are equal to one, i.e., $N_{0}=1$.
First, we have the following results.

\begin{Lem1}
Given the transmit power levels $P_{S}$ and $P_{R}$, the achievable throughput of the proposed scheme can be expressed as
\begin{small}
\begin{align}
C_{ADB}(P_{S},P_{R})
&=\frac{1}{2}{\rm min}\Bigg\{{\mathbb E}\left[{\rm log_{2}}\left(1+P_{S}\min\limits_{R_{i}\in{\mathcal R_{1}}}(|g_{i}|^{2})\right)\right],
\ \nonumber\\
&{\mathbb E}\left[{\rm log_{2}}\left(1+P_{R}\left(\sum\limits_{R_{i}\in{\mathcal R_{1}}}|h_{i}|\right)^{2}\right)\right]\Bigg\}
\ \nonumber\\
&+\frac{1}{2}{\rm min}\Bigg\{{\mathbb E}\left[{\rm log_{2}}\left(1+P_{S}\min\limits_{R_{i}\in{\mathcal R_{2}}}(|g_{i}|^{2})\right)\right],
\ \nonumber\\
&{\mathbb E}\left[{\rm log_{2}}\left(1+P_{R}\left(\sum\limits_{R_{i}\in{\mathcal R_{2}}}|h_{i}|\right)^{2}\right)\right]\Bigg\}.\label{throughput2}
\end{align}
\end{small}
\end{Lem1}

\emph{Proof:} Consider each group of relays, we know that the achievable throughput of the data flow passing through group ${\mathcal R_{1}}$ is given by \cite{4}
\begin{small}
\begin{align}
&C_{ADB_1}
\ \nonumber\\
&={\rm min}\Bigg\{\lim\limits_{N\rightarrow\infty}\frac{1}{N}\sum\limits_{t_{1}\in{\mathcal T_{1}}}\left[{\rm log_{2}}\left(1+P_{S}\min\limits_{R_{i}\in{\mathcal R_{1}}}\left(|g_{i}(t_{1})|^{2}\right)\right)\right],
\ \nonumber\\
&\lim\limits_{N\rightarrow\infty}\frac{1}{N}\sum\limits_{t_{2}\in{\mathcal T_{2}}}\left[{\rm log_{2}}\left(1+P_{R}\left(\sum\limits_{R_{i}\in{\mathcal R_{1}}}|h_{i}(t_{2})|\right)^{2}\right)\right]\Bigg\}
\ \nonumber\\
&=\frac{1}{2}{\rm min}\Bigg\{{\mathbb E}\left[{\rm log_{2}}\left(1+P_{S}\min\limits_{R_{i}\in{\mathcal R_{1}}}(|g_{i}|^{2})\right)\right],
\ \nonumber\\
&{\mathbb E}\left[{\rm log_{2}}\left(1+P_{R}\left(\sum\limits_{R_{i}\in{\mathcal R_{1}}}|h_{i}|\right)^{2}\right)\right]\Bigg\},
\end{align}
\end{small}
where $g_{i}(t_{1})$ and $h_{i}(t_{2})$ denote the channel coefficients of the $S-R_i$ and $R_i-D$ links in the $t_1$-th and $t_2$-th time slot, respectively, and the last equality comes from the fact that each transmission modes occupies half of the whole transmission time slots, i.e., $|{\mathcal T_{1}}|=|{\mathcal T_{2}}|=\frac{N}{2}$. The computation of $C_{ADB_2}$ is similar to that of $C_{ADB_1}$. Therefore, the achievable throughput of ADB can be calculated as $$C_{ADB}(P_{S},P_{R}) = C_{ADB_1} + C_{ADB_2}$$ as shown in (\ref{throughput2}).
\hfill$\square$

\begin{Rem}
The mode switching frequency $M$, i.e., the relays selected to receive data in every $M$ consecutive time slots, does not change the throughput.
It is obvious that as $M$ increases, the queue length increases and the average delay increases as well. Without loss of generality, we consider the case that $\mathcal{T}_{1}=\{1,3,5...\}$ while $\mathcal{T}_{2}=\{2,4,6...\}$ throughout this paper, i.e., the two different transmission modes alternates every time slot.
\end{Rem}

Denote
\begin{align}
&C_{11}={\mathbb E}\left[{\rm log_{2}}\left(1+P_{S}\min\limits_{R_{i}\in{\mathcal R_{1}}}(|g_{i}|^{2})\right)\right],\ \\
&C_{12}={\mathbb E}\left[{\rm log_{2}}\left(1+P_{R}\left(\sum\limits_{R_{i}\in{\mathcal R_{2}}}|h_{i}|\right)^{2}\right)\right],\ \\
&C_{21}={\mathbb E}\left[{\rm log_{2}}\left(1+P_{S}\min\limits_{R_{i}\in{\mathcal R_{2}}}(|g_{i}|^{2})\right)\right],\ \\
&C_{22}={\mathbb E}\left[{\rm log_{2}}\left(1+P_{R}\left(\sum\limits_{R_{i}\in{\mathcal R_{1}}}|h_{i}|\right)^{2}\right)\right].
\end{align}

\begin{Lem1}\label{prop:closed-form}
Given $P_{S}$ and $P_{R}$, the approximate closed-form expressions for the achievable throughput of ADB in Rayleigh fading channels are given by
\end{Lem1}
\begin{align}
\C_{ADB}&=\frac{1}{2}{\rm min}(\C_{11},\C_{22})+\frac{1}{2}{\rm min}(\C_{21},\C_{12}) \nonumber\\
&\hspace{-.5cm}=\left\{\begin{array}{ll}
\frac{1}{2}(\C_{11}+\C_{21}) & \text{if}\quad \C_{11}<\C_{22}\quad \text{and} \quad \C_{21}<\C_{12}, \\
\frac{1}{2}(\C_{11}+\C_{12}) & \text{if}\quad \C_{11}<\C_{22}\quad \text{and} \quad \C_{21}>\C_{12}, \\
\frac{1}{2}(\C_{22}+\C_{21}) & \text{if}\quad \C_{11}>\C_{22}\quad \text{and} \quad \C_{21}<\C_{12}, \\
\frac{1}{2}(\C_{22}+\C_{12}) & \text{otherwise}.\label{4cases}
\end{array}\right.
\end{align}
where
\begin{small}
\begin{align}
&\C_{11}=\frac{e^\frac{m}{2\sigma_{g}^{2}P_{S}}}{{\rm ln2}}E_{1}\left(\frac{m}{2\sigma_{g}^{2}P_{S}}\right),\ \\
&\C_{22}=\frac{1}{{\rm ln2}}\Bigg\{e^{\frac{1}{2b_{1}P_{R}m}}E_{1}\left(\frac{1}{2b_{1}P_{R}m}\right)\ \nonumber\\
&\times\left[\left(\sum\limits_{k=1}^{m-1}\frac{(-\frac{1}{P_{R}m})^{k}}{(2b_{1})^{k}\cdot{k!}}\right)+1\right]\ \nonumber\\
&+\sum\limits_{k=1}^{m-1}\frac{1}{(2b_{1})^{k}\cdot{k!}}\sum\limits_{s=1}^{k}(s-1)!\left(-\frac{1}{P_{R}m}\right)^{k-s}\left(\frac{1}{2b_{1}}\right)^{-s}\Bigg\},\ \\
&\C_{21}=\frac{e^\frac{L-m}{2\sigma_{g}^{2}P_{S}}}{{\rm ln2}}E_{1}\left(\frac{L-m}{2\sigma_{g}^{2}P_{S}}\right),
\end{align}
\end{small}
\begin{small}
\begin{align}
&\C_{12}=\frac{1}{{\rm ln2}}\Bigg\{e^{\frac{1}{2b_{2}P_{R}(L-m)}}E_{1}\left(\frac{1}{2b_{2}P_{R}(L-m)}\right)\ \nonumber\\
&\times\left[\left(\sum\limits_{k=1}^{L-m-1}\frac{\left(-\frac{1}{P_{R}(L-m)}\right)^{k}}{(2b_{2})^{k}\cdot{k!}}\right)+1\right]+\sum\limits_{k=1}^{L-m-1}\frac{1}{(2b_{2})^{k}\cdot{k!}} \nonumber\\
&\times\sum\limits_{s=1}^{k}(s-1)!\left(-\frac{1}{P_{R}(L-m)}\right)^{k-s}\left(\frac{1}{2b_{2}}\right)^{-s}\Bigg\},
\end{align}
\end{small}
where $$b_{1}=\frac{\sigma_{h}^{2}}{m}[(2m-1)!!]^{\frac{1}{m}}, $$ $$b_{2}=\frac{\sigma_{h}^{2}}{L-m}[(2(L-m)-1)!!]^{\frac{1}{L-m}},$$ and $(2m-1)!!=(2m-1)(2m-3)\cdot\cdot\cdot3\cdot1$, $E_{1}(x)=\int_{x}^\infty(e^{-t}/t)dt, x>0$ is the exponential integral function.

\emph{Proof:} Please see Appendix \ref{app:closed-form}.\hfill$\square$

Given the total power constraint SNR of the network, we can allocate the total power to the source and relays to achieve the best performance.

For ADB, the receiving and transmitting relays for each time slot have been determined before the system starts its normal operation. The source transmits in every time slot, while either $m$ relays in group ${\mathcal R_{1}}$ or $L-m$ relays in group ${\mathcal R_{2}}$ transmits in one time slot. Note also that $|{\mathcal T_{1}}|=|{\mathcal T_{2}}|=\frac{N}{2}$. Therefore, we should have $P_{S}+\frac{L}{2}P_{R} \leq {\rm SNR}$. Regarding SFD-MMRS, we allocate transmit power to the source and $L$ relays to enable each relay to be capable of being selected for transmission. Again, the sources works for all time slots. So we should have $P_{S}+LP_{R} \leq {\rm SNR}$. With CRS, similarly, we should allocate transmit energy to the source and $L$ relays, albeit the data transmission occupies two time slots. Therefore, we should have $\frac{1}{2}(P_{S}+LP_{R}) \leq {\rm SNR}$. With regard to DF, each relay must decode the common message transmitted by the source node and beamform their transmissions to the destination, obviously we need to allocate transmit energy to source and $L$ relays. It is also performed in two time slots, so we should have $\frac{1}{2}(P_{S}+LP_{R}) \leq {\rm SNR}$.

Consider the achievable throughput in (\ref{throughput2}), once given the total power SNR, it is obvious that when $P_{S}$ is small, the throughput is limited by the source-relay link. On the other hand, when $P_R$ is small, the relay-destination link will be the bottleneck of the system. Therefore, there is always an optimal power allocation that maximizes the achievable throughput.
\begin{Def}
The maximum achievable throughput of ADB is given by
\begin{align}
C_{max}=\max\limits_{P_{S}+\frac{L}{2}P_{R} \leq SNR}C_{ADB}(P_{S},P_{R}).
\end{align}
\end{Def}
Similarly, we can define the maximum achievable throughput for DF, CRS, and SFD-MMRS.

\section{Numerical Results}

In this section, we evaluate the proposed ADB scheme and compare it with that of CRS \cite{CRS}, SFD-MMRS \cite{mimick}, and DF \cite{DF}. We assume that $\sigma_{g}^{2}=\sigma_{h}^{2}=1$, unless specified otherwise.

\begin{figure}
    \centering
    \includegraphics[width=\figsize\textwidth]{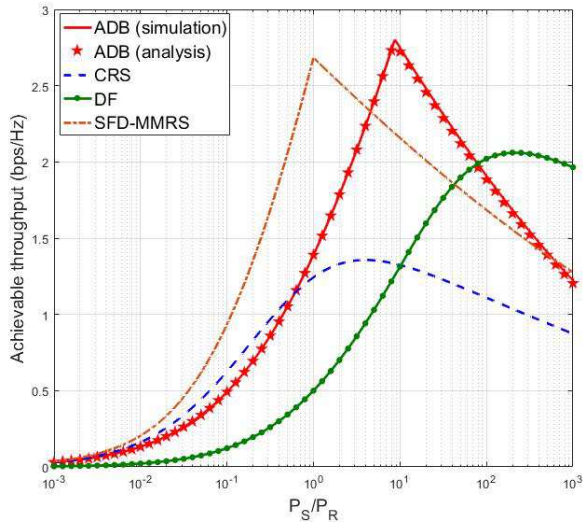}
    \caption{Achievable throughput versus $P_{S}/P_{R}$ for several relaying protocols.}
    \label{figure3}
\end{figure}
Fig. \ref{figure3} plots the achievable throughput versus $P_{S}/P_{R}$ for each scheme. We assume ${\rm SNR}=10$ dB, $L=4$ and $m=2$. We can find that the achievable throughput always has a peak value as $P_{S}/P_{R}$ varies, and the proposed scheme achieves the largest throughput. We also note that the analytical results obtained based on the derivation in Section III match the simulation results, which verifies the approximate closed-form expressions.

\begin{figure}
    \centering
    \includegraphics[width=\figsize\textwidth]{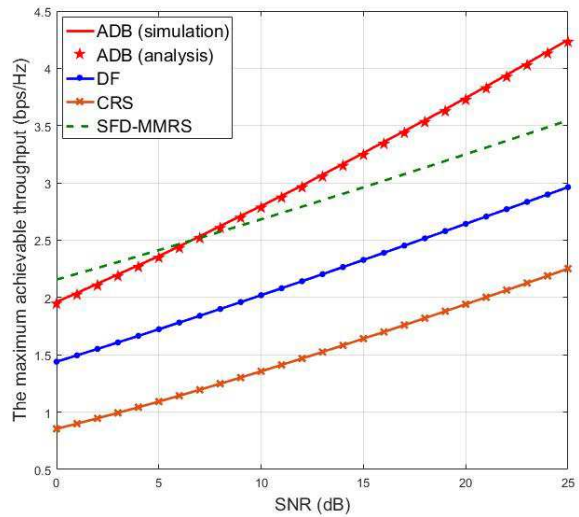}
    \caption{The maximum achievable throughput versus SNR for several relaying protocols.}
    \label{figure4}
\end{figure}
In Fig. \ref{figure4}, we compare the maximum achievable throughput of the proposed ADB scheme with that of two relay selection schemes and the traditional DF scheme as SNR varies. We assume $L=4$, and $m=2$. We can find that the proposed scheme achieves great performance gain especially in high SNR. We can see that the HD loss is recovered and a beamform gain can be achieved through power allocation.

\begin{figure}
    \centering
    \includegraphics[width=\figsize\textwidth]{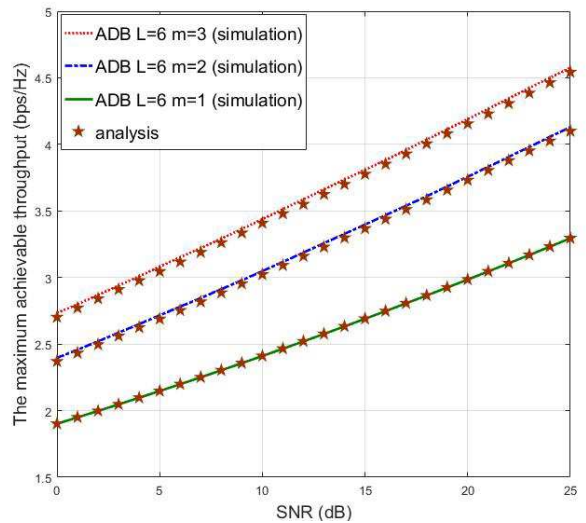}
    \caption{The maximum achievable throughput versus SNR for different grouping modes of the proposed scheme.}
    \label{figure5}
\end{figure}
In Fig. \ref{figure5}, we compare the maximum achievable throughput of the ADB scheme versus SNR for different $m$, i.e., different grouping modes. We assume $L=6$. It is interesting that, the symmetric allocation of relays achieves the best performance with the given setting. This is generally because that beamforming gain can be attained within each group.

\begin{figure}
    \centering
    \includegraphics[width=\figsize\textwidth]{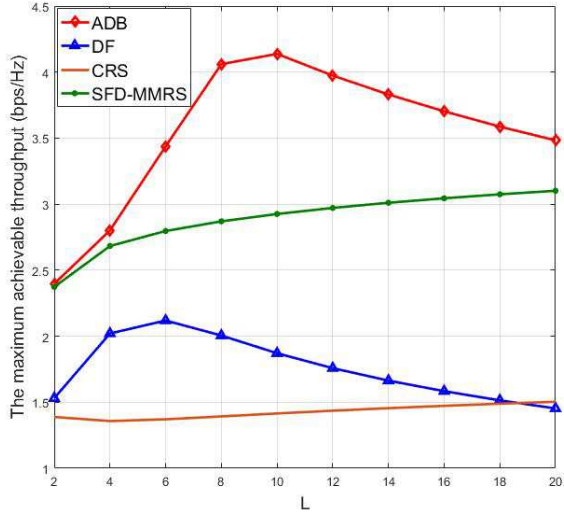}
    \caption{The maximum achievable throughput versus the number of relays for several relaying protocols.}
    \label{figure6}
\end{figure}
In Fig. \ref{figure6}, we plot the maximum achievable throughput of each scheme versus the number of relays for SNR = 10 dB. We assume $m=L/2$. We can find that the proposed scheme achieves the best performance in all cases. It is interesting that for a given SNR, the proposed scheme achieves the largest maximum throughput when $L=10$, and the DF strategy achieves the best throughput performance when $L=6$. This is probably due to the tradeoff between the reduction in the power allocated to relays and the increased possibility of selecting channels with better channel conditions as $L$ increases considering the total power constraint.

\section{Conclusion}
In this paper, we have proposed a new relay policy named ADB for buffer-aided multiple relay systems, in which the relays are divided into two groups, with one group receiving the signals transmitted by the source node while at the same time, the other group beamform the previously received data to the destination. In this policy, the relays used for reception and transmission are determined without the need of channel state information (CSI). We have obtained the closed-form expressions of the achievable throughput in Rayleigh fading channels. Through numerical results, we have found that the proposed scheme achieves significant improvement over the existing schemes in terms of the maximum achievable throughput. In addition, for a given total power constraint SNR, the proposed scheme achieves the best throughput performance at a mediate number of relays.

\appendix
\subsection{Proof of Proposition \ref{prop:closed-form}}\label{app:closed-form}
To compute the achievable throughput of ADB in (\ref{4cases}), we need to find $\C_{11}$, $\C_{12}$, $\C_{21}$, and $\C_{22}$.

\quad $Computation$ $of$ $\C_{11}$: In this case, we denote $z={\rm min}(g_{1}^{2},g_{2}^{2},...,g_{m}^{2})$. Therefore, to derive $\C_{11}$, we first compute the probability density function (PDF) of $z$.
The cumulative distribution function (CDF) of $z$ is given by
\begin{align}
F_{Z}(z)
&=P({\rm min}(g_{1}^{2},g_{2}^{2},...,g_{m}^{2})\leq z)\ \nonumber\\
&=1-P({\rm min}(g_{1}^{2},g_{2}^{2},...,g_{m}^{2})\geq z)\ \nonumber\\
&=1-P(g_{1}^{2}\geq z)P(g_{2}^{2}\geq z)...P(g_{m}^{2}\geq z)\ \nonumber\\
&=1-e^{-\frac{m}{2\sigma_{g}^{2}}z}.\label{zcdf}
\end{align}
Take the derivative of (\ref{zcdf}), the PDF of $z$ can be computed as $f_{Z}(z)=\frac{m}{2\sigma_{g}^{2}}e^{-\frac{m}{2\sigma_{g}^{2}}z}$. Then $\C_{11}$ can be obtained as
\begin{small}
\begin{align}
\C_{11}
&={\mathbb E}[{\rm log_{2}}(1+P_{S}z)]=\int_{0}^{\infty}{\rm log_{2}}(1+P_{S}z)f_{Z}({z})dz\ \nonumber\\
&=\int_{0}^{\infty}{\rm log_{2}}(1+P_{S}z)d\left(-e^{-\frac{m}{2\sigma_{g}^{2}}z}\right)\ \nonumber\\
&=-{\rm log_{2}}(1+P_{S}z)e^{-\frac{m}{2\sigma_{g}^{2}}z}|_{0}^{\infty}+\int_{0}^{\infty}e^{-\frac{m}{2\sigma_{g}^{2}}z}d\left({\rm log_{2}}(1+P_{S}z)\right)
\ \nonumber\\
&=\frac{1}{{\rm ln2}}\int_{0}^{\infty}\left(z+\frac{1}{P_{S}}\right)^{-1}e^{-\frac{m}{2\sigma_{g}^{2}}z}dz\ \nonumber\\
&=\frac{e^\frac{m}{2\sigma_{g}^{2}P_{S}}}{{\rm ln2}}E_{1}\left(\frac{m}{2\sigma_{g}^{2}P_{S}}\right),\label{C11}
\end{align}
\end{small}
where $E_{1}(x)=\int_{x}^\infty(e^{-t}/t)dt, x>0$ is the exponential integral function.
The computation of $\C_{21}$ is similar to that of $\C_{11}$, it is given by
\begin{align}
\C_{21}=\frac{e^\frac{L-m}{2\sigma_{g}^{2}P_{S}}}{{\rm ln2}}E_{1}\left(\frac{L-m}{2\sigma_{g}^{2}P_{S}}\right).\label{C21}
\end{align}
\quad $Computation$ $of$ $\C_{22}$: In this case, let $z=\sum\limits_{i=1}^{m}h_{i}$ be a sum of $m$ i.i.d. Rayleigh random variables (RV's).
Note that the distribution of an arbitrary sum of Rayleigh RV's is not exist in closed-form, as a result, numerical evaluations and approximations must be used \cite{notexist}.
A relatively simple and widely used small argument approximation (SAA) for the sum PDF was derived in \cite{pdf}. Then the SAA to the PDF of $z$ is
\begin{align}
f_{SAA}(t)=\frac{t^{(2m-1)}e^{-\frac{t^2}{2b}}}{2^{m-1}b^{m}(m-1)!},\ \nonumber\\
b=\frac{\sigma_{h}^{2}}{m}[(2m-1)!!]^{\frac{1}{m}},\label{fSAA}
\end{align}
where $(2m-1)!!=(2m-1)(2m-3)\cdot\cdot\cdot3\cdot1$ and $t=z/\sqrt{m}$ is the normalized argument.
Integration of (\ref{fSAA}) yields a SAA to the CDF of a Rayleigh sum given by
\begin{align}
F_{SAA}(t)=1-e^{-\frac{t^{2}}{2b}}\sum\limits_{k=0}^{m-1}\frac{\left(\frac{t^{2}}{2b}\right)^{k}}{k!}.
\end{align}
\allowdisplaybreaks[2]
Then the $\C_{22}$ can be computed as
\begin{align}
\C_{22}
&={\mathbb E}\left[{\rm log_{2}}\left(1+P_{R}(h_{1}+h_{2}+...+h_{m})^{2}\right)\right]\ \nonumber\\
&=\int_{0}^{\infty}{\rm log_{2}}\left(1+P_{R}(\sqrt{m}t)^{2}\right)f(t)dt\ \nonumber\\
&=\int_{0}^{\infty}{\rm log_{2}}(1+P_{R}mt^{2})d\left(F(t)-1\right)\ \nonumber\\
&={\rm log_{2}}(1+P_{R}mt^{2})(F(t)-1)|_{0}^{\infty}\ \nonumber\\
&-\int_{0}^{\infty}(F(t)-1)d\left({\rm log_{2}}(1+P_{R}mt^{2})\right)\ \nonumber\\
&=\int_{0}^{\infty}e^{-\frac{t^{2}}{2b}}\sum\limits_{k=0}^{m-1}\frac{(\frac{t^{2}}{2b})^{k}}{k!}d\left({\rm log_{2}}(1+P_{R}mt^{2})\right)\ \nonumber\\
&\xlongequal{x=t^{2}}\int_{0}^{\infty}e^{-\frac{x}{2b}}\sum\limits_{k=0}^{m-1}\frac{(\frac{x}{2b})^{k}}{k!}d\left({\rm log_{2}}(1+P_{R}mx)\right)\ \nonumber\\
&=\frac{1}{{\rm ln2}}\sum\limits_{k=1}^{m-1}\frac{1}{(2b)^{k}\cdot{k!}}\int_{0}^{\infty}\frac{x^{k}e^{-\frac{1}{2b}x}}{x+\frac{1}{P_{R}m}}dx\ \nonumber\\
&+\frac{1}{{\rm ln2}}\int_{0}^{\infty}\frac{e^{-\frac{1}{2b}x}}{x+\frac{1}{P_{R}m}}dx\ \nonumber\\
&=\frac{1}{{\rm ln2}}\Bigg\{e^{\frac{1}{2bP_{R}m}}E_{1}\left(\frac{1}{2bP_{R}m}\right)\left[\left(\sum\limits_{k=1}^{m-1}\frac{(-\frac{1}{P_{R}m})^{k}}{(2b)^{k}\cdot{k!}}\right)+1\right]\ \nonumber\\
&+\sum\limits_{k=1}^{m-1}\frac{1}{(2b)^{k}\cdot{k!}}\sum\limits_{s=1}^{k}(s-1)!\left(-\frac{1}{P_{R}m}\right)^{k-s}\left(\frac{1}{2b}\right)^{-s}\Bigg\},\label{C22}
\end{align}
where Eq. 3.353.5 in \cite{integral} is used to obtain the final equality.
The computation of $\C_{12}$ is similar to that of $\C_{22}$, it is given by
\begin{small}
\begin{align}
\C_{12}
&=\frac{1}{{\rm ln2}}\Bigg\{e^{\frac{1}{2bP_{R}(L-m)}}E_{1}\left(\frac{1}{2bP_{R}(L-m)}\right)\ \nonumber\\
&\times\left[\left(\sum\limits_{k=1}^{L-m-1}\frac{\left(-\frac{1}{P_{R}(L-m)}\right)^{k}}{(2b)^{k}\cdot{k!}}\right)+1\right]\ \nonumber\\
&\hspace{-.5cm}+\sum\limits_{k=1}^{L-m-1}\frac{1}{(2b)^{k}\cdot{k!}}\sum\limits_{s=1}^{k}(s-1)!\left(-\frac{1}{P_{R}(L-m)}\right)^{k-s}\left(\frac{1}{2b}\right)^{-s}\Bigg\},\label{C12}
\end{align}
\end{small}
where $b=\frac{\sigma_{h}^{2}}{L-m}[(2(L-m)-1)!!]^{\frac{1}{L-m}}$.
Finally, $\C_{ADB}$ is obtained by substituting (\ref{C11}), (\ref{C21}), (\ref{C22}), and (\ref{C12}) into (\ref{4cases}).

\balance


\begin{thebibliography}{1}

\bibitem{cooperative communication} J. N. Laneman, D. N. C. Tse, and G. W. Wornell, ``Cooperative diversity in wireless networks: efficient protocols and outage behavior,'' \emph{IEEE Trans. Inf. Theory}, vol. 50, no. 12, pp. 3062-3080, Dec. 2004.

\bibitem{DF} F. Parvaresh and R. H. Etkin, ``Using superposition codebooks and partial decode-and-forward in low-SNR parallel relay networks,'' \emph{IEEE Trans. Inf. Theory}, vol. 59, no. 3, pp. 1704-1723, Mar. 2013.

\bibitem{CRS} A. Bletsas, A. Khisti, D. Reed, and A. Lippman, ``A simple cooperative diversity method based on network path selection,'' \emph{IEEE J. Sel. Areas Commun.}, vol. 24, no. 3, pp. 659-672, Mar. 2006.

\bibitem{buffer1} N. Zlatanov, A. Ikhlef, T. Islam, and R. Schober, ``Buffer-aided cooperative communications: opportunities and challenges,'' \emph{IEEE Commun. Mag.}, vol. 52, no. 4, pp. 146-153, Apr. 2014.

\bibitem{buffer2} N. Zlatanov, R. Schober, and P. Popovski, ``Buffer-aided relaying with adaptive link selection,'' \emph{IEEE J. Sel. Areas Commun.}, vol. 31, no. 8, pp. 1530-1542, Aug. 2013.

\bibitem{max-max} A. Ikhlef, D. S. Michalopoulos, and Robert Schober, ``Max-max relay selection for relays with buffers,'' \emph{IEEE Trans. Wireless Commun.}, vol. 11, no. 3, pp. 1124-1135, Mar. 2012.

\bibitem{mimick} A. Ikhlef, J. Kim, and R. Schober, ``Mimicking full-duplex relaying using half-duplex relays with buffers,'' \emph{IEEE Trans. Veh. Technol.}, vol. 61, no. 7, pp. 3025-3037, Sep. 2012.

\bibitem{max-link} I. Krikidis, T. Charalambous, and J. S. Thompson, ``Buffer-aided relay selection for cooperative diversity systems without delay constraints,'' \emph{IEEE Trans. Wireless Commun.}, vol. 11, no. 5, pp. 1957-1967, May. 2012.

\bibitem{modified max-link} T. Charalambous, N. Nomikos, I. Krikidis, D. Vouyioukas, and M. Johansson, ``Modeling buffer-aided relay selection in networks with direct transmission capability.'' \emph{IEEE Commun. Lett.}, vol. 19, no. 4, pp. 649-652, Apr. 2015.

\bibitem{weight} W. Raza, N. Javaid, H. Nasir, N. Alrajeh, and N. Guizani, ``Buffer-aided relay selection with equal-weight links in cooperative wireless networks.'' \emph{IEEE Commun. Lett.}, vol. 22, no. 1, pp. 133-136, Jan. 2018.

\bibitem{2} M. Jain et al., ``Practical, real-time, full duplexing wireless,'' \emph{in Proc. ACM Mobile Comput. Netw. (MobiCom)}, Sep. 2011,
pp. 301¨C312.

\bibitem{3}  D. Bharadia, E. McMilin, and S. Katti, ``Full duplex radio,'' \emph{in Proc. ACM SIGCOMM Conf. Appl. Technol., Archit., Protocols Comput. Commun.}, Hong Kong, Aug. 2013, p. 1.

\bibitem{spectral efficient} B. Rankov and A. Wittneben, ``Spectral efficient protocols for half-duplex fading relay channels,'' \emph{IEEE J. Sel. Areas Commun.}, vol. 25, no. 2, pp. 379-389, Feb. 2007.

\bibitem{recover loss} Y. Fan, C. Wang, J. Thompson, and H. V. Poor, ``Recovering multiplexing loss through successive relaying using repetition coding,'' \emph{IEEE Trans. Wireless Commun.}, vol. 6, no. 12, pp. 4484-4493, Dec. 2007.

\bibitem{bypassing} D. S. Michalopoulos and G. K. Karagiannidis, ``Bypassing orthogonal relaying transmissions via spatial signal separation,'' \emph{IEEE Trans. Commun.}, vol.58, no. 10, pp. 3028-3038, Oct. 2010.


\bibitem{1} D. Qiao, ``Fixed versus selective scheduling for buffer-aided diamond relay systems under statistical delay constraints,'' \emph{IEEE Trans. on Commun.}, vol. 65, no. 7, pp. 2838-2851, July 2017.

\bibitem{frelay2} S. Wang, M. Xia, K. Huang, and Y. Wu, ``Wirelessly powered two-way communication with nonlinear energy harvesting model: rate regions under fixed and mobile relay,'' \emph{IEEE Trans. Wireless Commun.}, vol. 16, no. 12, pp. 8190-8204, Oct. 2017.

\bibitem{frelay3} C. B. Chae, T. Tang, R.W.Heath, Jr., and S.Cho, ``MIMO relaying with linear processing for multiuser transmission in fixed relay networks,'' \emph{IEEE Trans. Signal Process.}, vol. 56, no. 2, pp. 727-738, Feb. 2008.

\bibitem{4} B. Xia, Y. Fan, J. Thompson, H. Vincent Poor, ``Buffering in a three-node relay network,'' \emph{IEEE Trans. on Wireless Commun.}, vol. 7, no. 11, pp. 4492-4496, Nov. 2008.

\bibitem{notexist} J. Hu and N. O. Beaulieu, ``Accurate simple closed-form approximations to rayleigh sum distributions and densities,'' \emph{IEEE Commun. Lett.}, vol. 9, no. 2, pp. 109-111, Feb. 2005.

\bibitem{pdf} M. Schwartz, W. R. Bennett, and S. Stein, \emph{Communication Systems and Techniques.} New York: McGraw-Hill, 1966.

\bibitem{integral} I. S. Gradshteyn and I. M. Ryzhik, \emph{Tables of Integrals, Series, Products.}, 6th ed. San Diego, CA, USA: Academic, 2004.

\end{thebibliography}
\end{document}